\documentclass[twocolumn,08 pt,amsmath,amssymb,aps,fleqn]{revtex4-1}
\usepackage[hidelinks]{hyperref}
\usepackage[english]{babel}
\usepackage{mathtools}
\usepackage{tabularx}
\usepackage{multirow}
\usepackage{comment}
\usepackage{graphicx}
\usepackage{color}
\usepackage{float}
\usepackage{amsmath}
\usepackage{soul,cancel}
\begin{document}
\preprint{APS/123-QED}
\title{Unified Field Bosonization Technique for strongly inhomogenous Luttinger Liquids}
\author{Soundarya P$^{ 1}$, Venkata Suryanarayana M$^{ 1}$ and Joy Prakash Das$^{ 2*}$}
\affiliation{
	$^{\it 1}$Department of Physics, National Institute of Technology Tiruchirappalli, Tamil Nadu - 620015, India\\
	$^{\it 2}$Department of Physics, Assam Engineering College, Guwahati, Assam - 781013, India
		   }
\email{jpdas100@gmail.com}

\begin{abstract}
\begin{center}\bfseries Abstract\end{center}
We introduce the Unified Field Bosonization Technique (UFBT), a direct bosonization framework for strongly inhomogeneous one-dimensional Luttinger liquids (LLs) containing static impurities. UFBT incorporates impurity scattering through a symmetrized combination of bosonic phase fields and yields closed-form expressions for arbitrary N-point correlation functions for a broad class of static impurity potentials, including delta barriers, finite barriers and finite wells. The formalism requires neither renormalization-group analysis nor perturbative expansions, providing an analytical description of the inhomogeneous system at the bosonized level. The resulting correlation functions capture the leading singular contributions relevant to the present analysis. The technique is validated by recovering known limiting cases, showing agreement with the first-order perturbative expansion in the interaction strength, and demonstrating consistency with the Schwinger-Dyson equations. A key result is that the correlation function exponents remain independent of the impurity strength, while the impurity dependence is captured by the spatial structure and amplitudes of the correlation functions. The resulting correlation functions establish a foundation for analytical studies of transport, Friedel oscillations, and the local and dynamical density of states in strongly inhomogeneous Luttinger liquids.
\end{abstract}
\maketitle

\section{Introduction}
Impurities in strongly correlated one-dimensional (1D) systems can profoundly alter their low-energy properties, giving rise to rich and distinctive physical phenomena that have attracted considerable interest in condensed matter physics \cite{bahovadinov2022effects, gluza2022breaking, gotta2024dirac}. The enhanced role of interactions and quantum fluctuations in one dimension makes even a single impurity capable of drastically altering the physical properties of the system \cite{giamarchi2004quantum, furusaki2005kondo, das2019transport, kane1992transport}. Consequently, this problem has been approached from multiple perspectives over the years, leading to a wealth of theoretical insights and experimental progress \cite{kane1992transport, artemenko2005low, dinh2010tunneling, kainaris2018transmission, lo2019crossover}. Both analytical \cite{kane1992transport, grishin2004functional, matveev1993tunneling, das2018ponderous} and numerical techniques \cite{qin1996impurity, hamamoto2008numerical, freyn2011numerical,  moon1993resonant} have been extensively employed, complemented by experimental realizations in a variety of platforms \cite{ishii2003direct, auslaender2002tunneling, glazman1997new, bloch2008many, schwartz1998chain, dagotto1999experiments}.

A landmark contribution to this field was made by Kane and Fisher \cite{kane1992transport}, who demonstrated that impurities in 1D systems exhibit behavior that is qualitatively distinct from their higher-dimensional counterparts. Their work revealed the striking phenomena of effective “{\it cutting the chain}” for repulsive interactions and “{\it healing the chain}” for attractive interactions. These results underscored the nontrivial role played by impurities in interacting 1D systems and motivated subsequent efforts to obtain exact analytical expressions for correlation functions in the presence of both mutual interactions and disorder. Despite decades of research, deriving such exact results remains a formidable challenge. Nevertheless, sustained efforts over the past several decades have produced important partial results, often requiring compromises such as restricting the analysis to weak interactions, weak impurities, etc. Even so, these developments have significantly advanced our understanding of 1D correlated systems.

From an analytical standpoint, interactions in one dimension are commonly treated using the Fermi-Bose correspondence, wherein fermionic operators are expressed in terms of bosonic fields through the method of bosonization \cite{giamarchi2004quantum}. This framework enables a non-perturbative treatment of interactions and allows the calculation of $N$-point correlation functions for homogeneous 1D systems. However, within conventional bosonization schemes, impurities are typically incorporated only at a later stage using renormalization group techniques \cite{kane1992transport}. In contrast, the recently developed Non Chiral Bosonization technique (NCBT) introduces impurities directly at the level of the free fermion system and subsequently includes interactions \cite{das2018quantum}. While this approach is particularly well suited for treating impurity effects exactly, its analytical treatment is restricted to the most singular contributions to the two-point correlation functions, with higher-order correlation functions becoming considerably more formidable to obtain.

Several other analytical studies have addressed special limits of the impurity problem. Eggert {\it et al.} \cite{eggert1996boundary} obtained correlation functions for a Luttinger liquid with arbitrary interaction strength in the presence of open boundaries, which effectively correspond to infinitely strong impurities that completely suppress tunneling. On the other hand, Matveev {\it et al.} \cite{matveev1993tunneling} considered impurities of arbitrary strength but restricted the analysis to weak inter-particle interactions. Effective field theories of Luttinger liquids \cite{haldane1981luttinger, von1998bosonization} are inherently restricted to the low-energy sector near the Fermi surface. These constraints collectively highlight the limitations of purely analytical approaches and motivate the use of numerical methods.

On the numerical side, the density matrix renormalization group (DMRG) has become a powerful and reliable tool for studying interacting 1D systems \cite{white1992density, schollwock2005density}. Early DMRG studies by Qin {\it et al.} examined impurities in Luttinger liquids and verified key predictions of conformal field theory \cite{qin1996impurity, qin1997impurity}, while Schollw$\ddot{\text{o}}$ck {\it et al.} showed that bosonization predictions are strictly recovered only in the low-energy limit \cite{schollwock2002dmrg}. More recently, dynamical DMRG has been used to investigate conductance through single and double barriers, reproducing key field-theoretical predictions \cite{bischoff2017density}, while DMRG-based studies of weak links have explored the crossover of correlation functions from short- to long-distance regimes near an impurity \cite{lo2019crossover}.

In this work, we present, for the first time, a new bosonization framework, which we refer to as the Unified Field Bosonization Technique (UFBT). The method provides a unified analytical treatment of Luttinger liquids in the presence of arbitrary impurity and interaction strengths and enables the derivation of general $N$-point correlation functions without resorting to renormalization group techniques. Within this framework, the exponents of the correlation functions are independent of the strength of the impurity, while the impurity dependence is captured through the spatial structure and amplitudes of the correlation functions. Using UFBT, we explicitly derive the two-point and four-point correlation functions of a Luttinger liquid containing a cluster of impurities, demonstrating the applicability of the framework to multiple-impurity configurations. Within the present treatment, the correlation functions obtained from the quadratic cumulant capture their leading singular contributions. More broadly, UFBT provides a unified analytical framework for obtaining correlation functions in inhomogeneous interacting 1D systems, beyond the restricted limits accessible through existing analytical approaches.

\section{System Description}
Consider a one-dimensional (1D) system of interacting electrons in the presence of a cluster of impurities localized around an origin. At low energies, the clean interacting system is universally described by the Luttinger liquid framework, which captures the profound effects of electron-electron interactions in one dimension \cite{haldane1981luttinger, voit1995one, giamarchi2004quantum}. We consider the generic Hamiltonian
\begin{equation}
\begin{split}
H =\int_{-\infty}^{\infty} &dx \, \psi^\dagger(x) \left( -\frac{1}{2m} \partial_x^2 + V(x) \right) \psi(x) \\
&+ \frac{1}{2} \int_{-\infty}^{\infty} dx \int_{-\infty}^{\infty} dx' \, v(x - x') \, \rho(x) \rho(x')
\end{split}
\end{equation}
where $\rho(x)=\psi^\dagger(x)\psi(x)$ and $v(x-x')$ describes the electron-electron interaction. We restrict the interaction to forward scattering and assume it to be short ranged \cite{haldane1981luttinger,giamarchi2004quantum}. In Fourier space, $v(x - x') = \frac{1}{L} \sum_q v_q \exp[-iq(x - x')]$, where $v_q = 0$ if $|q| > \Lambda$ for some fixed bandwidth $\Lambda \ll k_F$, and $v_q = v_0$ is a constant otherwise \cite{das2019conductance}, with $L$ being the size of the system. 

The potential term $V(x)$ describes a localized cluster of impurities confined to a finite region of width $w$ around the origin, with an arbitrary configuration of barriers and wells. The corresponding reflection and transmission amplitudes can, in principle, be obtained from standard single-particle scattering theory and contain the detailed information about the impurity configuration. Since the subsequent derivation does not require their explicit forms, the amplitudes are retained in general form and denoted simply by $R$ and $T$ respectively (such that $|R|^2+|T|^2=1$). Thus, at low energies, the impurity cluster is characterized by the amplitudes $R$ and $T$, and the resulting correlation functions are expressed in terms of these quantities. The expressions thus obtained are therefore applicable to any specific impurity configuration upon substituting the explicit expressions of the corresponding reflection and transmission amplitudes calculated from $V(x)$.

To make analytical progress while retaining the essential scattering physics of the impurity, we consider the Random Phase Approximation (RPA) limit \cite{stone1994bosonization, aristov2009conductance}. This limit is obtained by sending the Fermi momentum and fermion mass to infinity ($k_F,m\rightarrow\infty$) while keeping the Fermi velocity finite ($v_F=\frac{k_F}{m}<\infty$), where we set $\hbar=1$. In this limit, the single-particle dispersion about either Fermi point becomes strictly linear for finite momentum measured relative to the corresponding Fermi point, $E=E_F+p v_F$, where $p$ denotes the momentum measured from the Fermi point. The scaling of the impurity potential must also be specified in this limit. In particular, if $w$ denotes the spatial extent of the impurity cluster, we keep the dimensionless quantity $k_F w$ finite as $k_F\rightarrow\infty$. Thus, although the Fermi wavelength vanishes in the RPA limit, the spatial extent of the impurity remains finite when measured in units of the Fermi wavelength. Similarly the heights and depths of the various barriers are assumed to be in fixed ratios with the Fermi energy $ E_F = \frac{1}{2} m v_F^2 $ even as $ m \rightarrow \infty $ with $ v_F < \infty $. This prescription ensures that the impurity retains nontrivial scattering properties in the RPA limit.

\section{Formalism of the Unified Field Bosonization Technique}
The unified field bosonization technique (UFBT) provides a general framework for evaluating correlation functions in strongly inhomogeneous Luttinger liquids. Its central ingredient is a unified Fermi-Bose correspondence that incorporates the spatial inhomogeneity of the system directly into the bosonized fermionic fields, while retaining the underlying field-theoretic structure required for the systematic evaluation of correlation functions. The evaluation of the correlation functions within this framework proceeds through a sequence of well-defined steps. Firstly, the single-particle two-point functions are first obtained for the noninteracting system in the RPA limit in the presence of the localized impurity cluster, from which the slow part of the density-density correlation function (DDCF) is determined. The fermionic two-point functions are then expressed in terms of these density fields through the Fermi-Bose correspondence, and the resulting DDCF is subsequently generalized to include the forward-scattering interactions. Finally, the interacting density fields are substituted into the bosonized fermionic correlation functions to obtain the interacting Green functions. These steps constitute the general procedure for evaluating the correlation functions and have been described in detail in the earlier formulation \cite{das2018quantum}; they are therefore only briefly reviewed here. The central focus of this section is the unified Fermi--Bose correspondence and its implementation, which constitute the defining elements of UFBT and provide a general framework for treating correlation functions in strongly inhomogeneous Luttinger liquids.
\subsection{Free-Fermion Two-Point Green's Functions}
 The full two-point Green function of the system for a parabolic dispersion model can be denoted by $\langle T \psi(x,\sigma,t)\psi^\dagger(x',\sigma',t') \rangle$, where $x,x'$ denote position, $t,t'$ time, and $\sigma,\sigma'$ spin, with the time-ordering operator $T$ determining whether the particle or hole Green function is considered. The asymptotic or RPA Green function $\langle T \psi_{\nu}(x,\sigma,t)\psi_{\nu'}^\dagger(x',\sigma',t') \rangle$ is obtained by coarse-graining the full Green function over spatial and temporal scales set by the Fermi wavelength ($\lambda_F=2\pi/k_F$) and Fermi time ($T_F=2\pi/E_F$) respectively, followed by taking the ($m\to\infty$) limit at fixed Fermi velocity ($v_F$). Here, $\nu=R,L$ denote the right- and left-moving Fermi points, respectively, with $R\equiv +1$ and $L\equiv -1$.

In the absence of electron-electron interactions, the two point RPA Green function has the following form (at zero temperature):
\begin{equation}
\begin{aligned}
\langle T \psi_\nu(x,\sigma,t)&\psi_{\nu'}^\dagger(x',\sigma',t') \rangle_0 \\
=& \delta_{\sigma,\sigma'} \sum_{\gamma,\gamma' = \pm 1} \frac{\theta(\gamma x)\theta(\gamma' x') \, g_{\gamma,\gamma'}(\nu,\nu')}{(\nu x - \nu' x') - v_F(t-t')}
\label{greenfunction0}
\end{aligned}
\end{equation}
where $\theta(x)$ is the Heaviside step function, and $\gamma,\gamma'=\pm1$ indicate the sides of the impurity (situated at the origin) on which the observation point $x$ and $x'$ lies, with $\gamma=+1$ for the point to the right and $\gamma=-1$ for the point to the left of the impurity. The quantity $g_{\gamma_1,\gamma_2}(\nu_1,\nu_2)$ incorporates the corresponding reflection ($R$) and transmission ($T$) amplitudes and is explicitly given by the following expression (here $\delta_{a,b} = 1$ when $a=b$ and $0$ otherwise is the Kronecker delta function).
\small
\begin{equation}
\begin{aligned}
g_{\gamma,\gamma'}(\nu,\nu') = \frac{i}{2\pi} \bigg[[ \delta_{\nu,\nu'}\delta_{\gamma,\gamma'}& + (T \delta_{\nu,\nu'} + R \delta_{\nu,-\nu'})\delta_{\gamma,\nu}\delta_{\gamma',-\nu'}\\
 +& (T^* \delta_{\nu,\nu'} + R^* \delta_{\nu,-\nu'})\delta_{\gamma,-\nu}\delta_{\gamma',\nu'} \bigg]
\label{gmunu}
\end{aligned}
\end{equation}\normalsize
While the non-interacting propagator in equation \ref{greenfunction0} exhibits a simple linear denominator, switching on forward-scattering interactions in the Luttinger liquid regime modifies this term to $[(\nu x - \nu' x') - v_F(t-t')]^g$. Calculating these correlation function exponents $g$ in presence of both impurity and mutual interactions, each of arbitrary strengths, forms a primary objective of this work. The extension to finite temperature follows directly by replacing the zero-temperature factor $1/X$ according to $\frac{1}{X}\longrightarrow \frac{\pi}{\beta v_F}\operatorname{csch}\left(\frac{\pi X}{\beta v_F}\right),$ where $X\equiv [(\nu x-\nu' x')-v_F(t-t')]$ and $\beta$ denotes the inverse temperature.

\subsection{Density Density Correlation functions}
The Density Density correlation function $\langle T \rho(x,\sigma,t)\rho(x',\sigma',t') \rangle_0$ can be expressed in terms of the two point functions using the Wick's theorem. Subtracting the average density terms (so that this is really the deviation), the density density correlation functions can be redefined as follows.\small
\begin{equation}
\begin{aligned}
\big<T\tilde{\rho}&(x,\sigma,t)\tilde{\rho}(x',\sigma',t')\big>\\
=&\mbox{ }\mbox{ }\mbox{ }\big<T\rho(x,\sigma,t)\rho(x',\sigma',t')\big> - \mbox{ }\big<\rho(x,\sigma,t)\big>\big<\rho(x',\sigma',t')\big>\\
=&-\big<T\psi(x,\sigma,t)\psi^{\dagger}(x',\sigma',t')\big>\big<T\psi(x',\sigma',t')\psi^{\dagger}(x,\sigma,t)\big>\\
\label{tilderho}
\end{aligned}
\end{equation}\normalsize
In the RPA framework, the density is decomposed into slowly varying and rapidly oscillating components, denoted by $\rho_s$ and $\rho_f$, respectively. The density-density correlation function (DDCF) consequently contains contributions from both components. Since only the slowly varying component is relevant for the present analysis, the DDCF in the absence of mutual interactions is obtained by retaining the contribution from $\rho_s$, as follows.
\begin{equation}
\begin{aligned}
\langle T \mbox{    }&\rho_s(x,t)\rho_s(x^{'},t^{'})\rangle_0\\
&=-\sum_{ \substack{\gamma,\gamma^{'} \\= \pm 1}}\hspace{0.2 cm}\sum_{ \substack{ \nu,\nu^{'}\\=\pm 1}} \frac{ |g_{\gamma,\gamma^{'}}(\nu,\nu^{'})|^2 \mbox{  } \theta(\gamma x) \theta(\gamma^{'} x^{'})
 }{ [(\nu x - \nu^{'} x^{'}) - v_F (t-t^{'}) ]^2 }
\label{INPUT3}
\end{aligned}
\end{equation}
\normalsize
where $ g_{\gamma,\gamma'} (\nu,\nu') $ are given in equation {\ref{gmunu}}.

\subsection{Bosonization via Fermi Bose Correspondence}
The fermionic field can be decomposed into right- and left-moving components as
$
\psi(x)=e^{ik_Fx}\psi_R(x)+e^{-ik_Fx}\psi_L(x).
$
Bosonization expresses the fermionic field in terms of the density and current. Using the continuity equation, the current can be eliminated in favor of $\rho$ and $\partial_t\rho$, allowing the single-particle Green functions to be represented through density correlations. In the standard $g$-ology formulation ($\nu=\pm1$) \cite{giamarchi2004quantum} ,\normalsize
\begin{equation}
\begin{aligned}
\hspace{0.9 in}
\psi_\nu(x,\sigma,t)\sim e^{i\theta_\nu(x,\sigma,t)}
\end{aligned}
\end{equation}
with the local phase given by the formula ($\nu=\pm1$),
\small
\begin{equation}
\begin{aligned}
\theta_{\nu}(x,\sigma,t) = \pi \int^{x}_{sgn(x)\infty}& dy \bigg( \nu  \mbox{  } \rho_s(y,\sigma,t) \\
&- \int^{y}_{sgn(y)\infty} dy^{'} \mbox{ }\partial_{v_F t }  \mbox{ }\rho_s(y^{'},\sigma,t) \bigg)
\end{aligned}
\end{equation}\normalsize
\normalsize
The above Fermi-Bose correspondence is formulated for nearly translationally invariant systems and half-line geometries and is therefore not directly applicable to the strongly inhomogeneous systems with arbitrary impurity strengths considered here. A new Fermi-Bose correspondence is consequently required, which constitutes the  central element of the Unified Field Bosonization technique. The new prescription modifies the dependence of the fermionic field on the bosonized phase field $\theta_\nu$, constructed from the density and current, by incorporating contributions from both sides of the inhomogeneous region through the combination $\theta_{\nu}(x)+\theta_{\nu}(-x)$. This provides a unified representation of the fermionic field across the inhomogeneity and serves as the basis for calculating its correlation functions.
\begin{equation}
\begin{aligned}
\hspace{0.5 in}
\psi_\nu(x,\sigma,t)\sim e^{\frac{\mbox{ }i}{\sqrt{2}}\left(\theta_\nu(x,\sigma,t)+\theta_\nu(-x,\sigma,t)\right)}
\label{ucbt}
\end{aligned}
\end{equation}
The symbol `$\sim$' emphasizes that this relation is a prescription for generating correlation functions rather than an exact operator identity. In particular, bosonization does not determine the model-dependent prefactors appearing in the Green functions. These prefactors are therefore fixed by matching the bosonized results to the corresponding non-interacting Green functions obtained using Fermi algebra. Since equation (\ref{ucbt}) is used only as a mnemonic for constructing the $N$-point functions, the explicit introduction of Klein factors, as required in the conventional operator formulation, is unnecessary.
The essential new feature of equation (\ref{ucbt}) is the contribution from $\theta_\nu(-x,\sigma,t)$. This term incorporates the effect of backscattering from the external inhomogeneity directly into the Fermi-Bose correspondence. Consequently, when the resulting expression is used to construct the $N$-point functions within the RPA framework, it reproduces the expected trivial exponents in the non-interacting limit, providing an essential consistency check of the correspondence.
While computing the Green’s functions using this prescription, the Gaussian approximation viz. $< e^Ae^B >= e^{\frac{1}{2} <A^2>}e^{\frac{1}{2} <B^2>}e^{\frac{1}{2} <AB>}$ is invoked. Introducing the notation $\Theta_{\nu}(x,\sigma,t) =(1/\sqrt{2}) ( \theta_\nu(x,\sigma,t)+\theta_\nu(-x,\sigma,t))$, the bosonized version of the two point functions can be written as follows.
\begin{equation}
\begin{aligned}
<\psi_{\nu}(x,\sigma,t)\psi_{\nu'}^{\dagger}(x',&\sigma,t') \mbox{ }\sim \mbox{ }<e^{i\Theta_{\nu}(x,\sigma,t)}e^{-i\Theta_{\nu'}(x',\sigma,t')}>\\
\sim &\mbox{ }e^{\frac{1}{2}<(i\Theta_{\nu}(x,\sigma,t))^2>}e^{\frac{1}{2}<(-i\Theta_{\nu'}(x',\sigma,t'))^2>}\\
&e^{<(i\Theta_{\nu}(x,\sigma,t))(-i\Theta_{\nu'}(x',\sigma,t'))>}
\label{bch}
\end{aligned}
\end{equation}
while the first two terms (of the type $<\Theta_{\nu}>^2$) become independent of time after proper calculation and are incorporated into the pre-factors. 
The third term in the above equation will lead to the correct term corresponding to the Green function upon substituting the explicit expression for the DDCF given in equation \ref{INPUT3}.
\begin{equation}
\begin{aligned}
e^{<\Theta_{\nu}(x,\sigma,t)\Theta_{\nu'}(x',\sigma,t')>} = \frac{1}{(\nu x-\nu'x')-v_F(t-t')}
\end{aligned}
\end{equation}

\subsection{Incorportation of forward scattering interactions}
Again in the spirit of the RPA, the density density correlation functions given in equation (\ref{INPUT3}) are modified to account for mutual interactions, yielding the following expressions \cite{danny2020density} (Note: $\rho_h(x,t) =  \rho_s(x,\uparrow,t) + \rho_s(x,\downarrow,t)  $ is the ``holon" density and $ \rho_{n}(x,t) =  \rho_s(x,\uparrow,t) - \rho_{s}(x,\downarrow,t)  $ is the ``spinon" density and $ a = h $ for holon and $ a = n $ for spinon) \cite{das2018quantum}
\footnotesize
\begin{equation}
\begin{aligned}
< T\mbox{   } \rho_s(x_1,\sigma_1,t_1)\rho_s(x_2,&\sigma_2,t_2)> =  \frac{1}{4}
\Big(< T\mbox{   } \rho_h(x_1,t_1)\rho_h(x_2,t_2)> \\
&+\sigma_1\sigma_2< T\mbox{   } \rho_n(x_1,t_1)\rho_n(x_2,t_2)>\Big)\\
\label{intddcf1}
\end{aligned}
\end{equation}
\normalsize
 where
\footnotesize
\begin{equation}
\begin{aligned}
\langle T\mbox{   } \rho_a(x_1,t_1)\rho_a(x_2,t_2)\rangle  = \frac{v_F  }{ 2\pi^2 v_a } \mbox{   } &\sum_{  \nu = \pm 1 }\bigg (   \frac{-1}{ ( x_1-x_2 + \nu v_a(t_1-t_2) )^2 }\\
&	-  \frac{\frac{v_F }{v_a}  \mbox{    } \text{sgn}(x_1) \text{sgn}(x_2)\mbox{   }Z_a}{  ( | x_1|+|x_2 | + \nu v_a(t_1-t_2) )^2 }
\bigg)
\label{intddcf2}
\end{aligned}
\end{equation}\normalsize
where $ a = n$ (spinon) or  h (holon) and,
\begin{equation}
\hspace{1 cm}
 Z_a = \frac{ |R|^2 }{    \bigg( 1 - \delta_{a,h} \frac{(v_h-v_F)}{ v_h }
 |R|^2   \bigg) }
 \label{Za}
\end{equation}
\normalsize
Here the spinon velocity is just the Fermi velocity since it is the total density that couples to the short-range potential:
 $ v_n = v_F $, but the holon velocity is modified by interactions, \scriptsize $ v_h = \sqrt{v_F^2+2v_F v_0/\pi} $ \normalsize where the interaction between fermions is the two-body short-range forward scattering potential which just means the potential between two particles at $ x $ and $ x^{'} $ is \footnotesize  $ v(x-x^{'}) = \frac{1}{L}\sum_{|q| < \Lambda }v_0 \mbox{ } \exp{[ -i q (x-x^{'})] } $, \normalsize   where $ \Lambda $ is held fixed as the RPA limit is taken. The holon and spinon sectors are decoupled, as reflected by $ \langle T \mbox{ }\rho_n(x_1,t_1)\rho_h(x_2,t_2)\rangle \equiv 0 $.  Furthermore, an expansion of equation (\ref{intddcf2}) in powers of $v_0$ can be shown to reproduce the corresponding series obtained from standard perturbation theory, provided only the most singular terms are retained \cite{das2018quantum}. The presence of two distinct velocities is a signature of spin-charge separation, a characteristic feature of one-dimensional interacting systems.

\subsection{Interacting DDCF in Bosonized Green's functions}
Once the density-density correlation functions (DDCFs) in the presence of mutual interactions have been obtained, the remaining step is to substitute the interaction-modified DDCFs given by equations (\ref{intddcf1}) and (\ref{intddcf2}) into the bosonized Green's functions given in equation (\ref{bch}). This substitution yields the dynamical part of the Green's functions for the system under consideration. Upon incorporating the prefactors determined by direct comparison with the corresponding non-interacting Green's functions, the complete Green's functions are obtained.

It is important to note that the space-time factors of the form $[(x_1-x_2)-v_a(t_1-t_2)]$ appearing in the correlation functions acquire non-trivial exponents in the presence of mutual interactions. These exponents depend explicitly on the strength of the interactions and characterize the modification of the correlation functions induced by the interactions. Determining and systematically listing these interaction-dependent exponents for the different correlation functions constitutes a central objective of the present work.

\section{Results and Discussion}
The Unified Field Bosonization Technique (UFBT) provides closed-form expressions for the leading singular contributions to the interacting single-particle Green's functions of a Luttinger liquid in the presence of a localized impurity cluster near the origin. Using the symmetrized bosonic-field correspondence in equation (\ref{ucbt}) together with the interacting density-density correlation function given by equations (\ref{intddcf1}) and (\ref{intddcf2}), the two-point Green's functions can be obtained directly. In the expressions below, the primary quantities of interest are the exponents of the factors of the form $[(\nu_1x_1-\nu_2x_2)-v(t_1-t_2)]^g$, since these exponents characterize the universal long-distance and long-time behaviour, whereas the associated coefficients are generally non-universal and depend on the impurity potential, short-distance cutoffs, and other microscopic details. The notation $X_i\equiv(x_i,\sigma_i,t_i)$ is used throughout. Since a translationally non-invariant system can contain cutoff-dependent, spatially varying time-independent factors, the symbol $A[X_1,X_2]\sim B[X_1,X_2]$ is used to denote weak equality, defined through $\partial_{t_1}\ln A=\partial_{t_1}\ln B$, provided the quantities do not vanish identically. This convention suppresses singular cutoff-dependent prefactors that are not relevant to the extraction of the temporal power-law exponents. Such prefactors arise naturally in the Gaussian evaluation of exponential bosonic operators, $\langle e^Ae^B\rangle=e^{\frac12\langle A^2\rangle}e^{\frac12\langle B^2\rangle}e^{\langle AB\rangle}$, where the first two factors can be spatially inhomogeneous but time independent. Retaining the quadratic cumulant yields the leading singular contribution to the Green function, with higher connected cumulants contributing at subleading singular order. Although these factors do not affect the power law exponents, they are essential when quantities such as the tunneling conductance and local dynamical density of states are evaluated. With $\tau_{12}\equiv t_1-t_2$, the resulting interacting Green's functions are given below, followed by a discussion about their limiting behaviour and a non-perturbative validation through the Schwinger--Dyson equation.
\scriptsize
\begin{equation*}
\begin{aligned}
\Big\langle T\psi&_{R}(X_1)\psi_{R}^{\dagger}(X_2)\Big\rangle \sim  \sum_{\substack{\gamma,\gamma'\\=\pm1}}
\frac{ \delta_{\sigma,\sigma'}\,\theta(\gamma x)\theta(\gamma' x') \, g_{\gamma,\gamma'}(1,1)}{(x_1-x_2 -v_h \tau_{12})^{P} (-x_1+x_2 -v_h \tau_{12})^{Q}} \\
\times&\frac{(4x_1x_2)^{X}}{ (x_1+x_2 -v_h \tau_{12})^{X} (-x_1-x_2 -v_h \tau_{12})^{X} (x_1-x_2 -v_F \tau_{12})^{0.5}}\\
\end{aligned}
\end{equation*}
\begin{equation}
\begin{aligned}
\Big\langle T\mbox{  }\psi&_{L}(X_1)\psi_{L}^{\dagger}(X_2)\Big\rangle \sim  \sum_{\substack{\gamma,\gamma'\\=\pm1}}
\frac{\delta_{\sigma,\sigma'}\,\theta(\gamma x)\theta(\gamma' x') \, g_{\gamma,\gamma'}(-1,-1)}{(x_1-x_2 -v_h \tau_{12})^{Q} (-x_1+x_2 -v_h \tau_{12})^{P}} \\
\times&\frac{(4x_1x_2)^{X}}{ (x_1+x_2 -v_h \tau_{12})^{X} (-x_1-x_2 -v_h \tau_{12})^{X}(-x_1+x_2 -v_F \tau_{12})^{0.5}}\\
\Big\langle T\mbox{  }\psi&_{R}(X_1)\psi_{L}^{\dagger}(X_2)\Big\rangle \sim  \sum_{\substack{\gamma,\gamma'\\=\pm1}}
\frac{\delta_{\sigma,\sigma'}\,\theta(\gamma x)\theta(\gamma' x') \, g_{\gamma,\gamma'}(1,-1)}{(x_1-x_2 -v_h \tau_{12})^{X} (-x_1+x_2 -v_h \tau_{12})^{X}} \\
\times&\frac{(4x_1x_2)^{X}}{ (x_1+x_2 -v_h \tau_{12})^{P} (-x_1-x_2 -v_h \tau_{12})^{Q} (x_1+x_2 -v_F \tau_{12})^{0.5}}\\
\Big\langle T\mbox{  }\psi&_{L}(X_1)\psi_{R}^{\dagger}(X_2)\Big\rangle \sim  \sum_{\substack{\gamma,\gamma'\\=\pm1}}
\frac{\delta_{\sigma,\sigma'}\,\theta(\gamma x)\theta(\gamma' x') \, g_{\gamma,\gamma'}(-1,1)}{(x_1-x_2 -v_h \tau_{12})^{X} (-x_1+x_2 -v_h \tau_{12})^{X}} \\
\times&\frac{(4x_1x_2)^{X}}{ (x_1+x_2 -v_h \tau_{12})^{Q} (-x_1-x_2 -v_h \tau_{12})^{P}(-x_1-x_2 -v_F \tau_{12})^{0.5}}\\
\label{GFall}
\end{aligned}
\end{equation}
\normalsize
Here, $v_h$ and $v_F$ denote the holon and spinon velocities, respectively, which characterize spin-charge separation and were discussed in the preceding section. The explicit expressions for the correlation-function exponents, denoted by the $g$'s in  equations \ref{GFall}, are given by
\begin{equation}
P = \frac{(v_h+v_F)^2}{8v_hv_F}
\ ;\quad
Q = \frac{(v_h-v_F)^2}{8v_hv_F}
\ ;\quad
X = \frac{v_h^2-v_F^2}{8v_hv_F}.
\label{exponents}
\end{equation}
A central observation is that all three exponents, $P$, $Q$, and $X$, are independent of the impurity strength, as characterized by the reflection amplitudes. They depend only on the bulk velocities $v_h$ and $v_F$, which are determined by the interaction strength $v_0$. Consequently, the algebraic decay of the correlation functions in the presence of a localized impurity cluster is governed solely by the bulk Luttinger-liquid parameters and remains independent of the impurity strength. In terms of the Luttinger parameter $K=v_F/v_h$, these exponents therefore retain a purely bulk origin. This result shows that, within UFBT, the correlation-function exponents are determined entirely by the bulk interaction parameters and remain independent of the impurity strength. This differs from the Non-Chiral Bosonization approach \cite{das2019nonchiral}, in which the exponents of the full Green's functions depend explicitly on the impurity parameters. A key advantage of UFBT is that these exponents are obtained directly within the bosonization framework, without requiring an RG treatment, as is commonly employed in g-ology approaches, and without restricting the impurity or interaction strengths.

\begin{figure}[h]
\begin{center}
\includegraphics[scale=0.35]{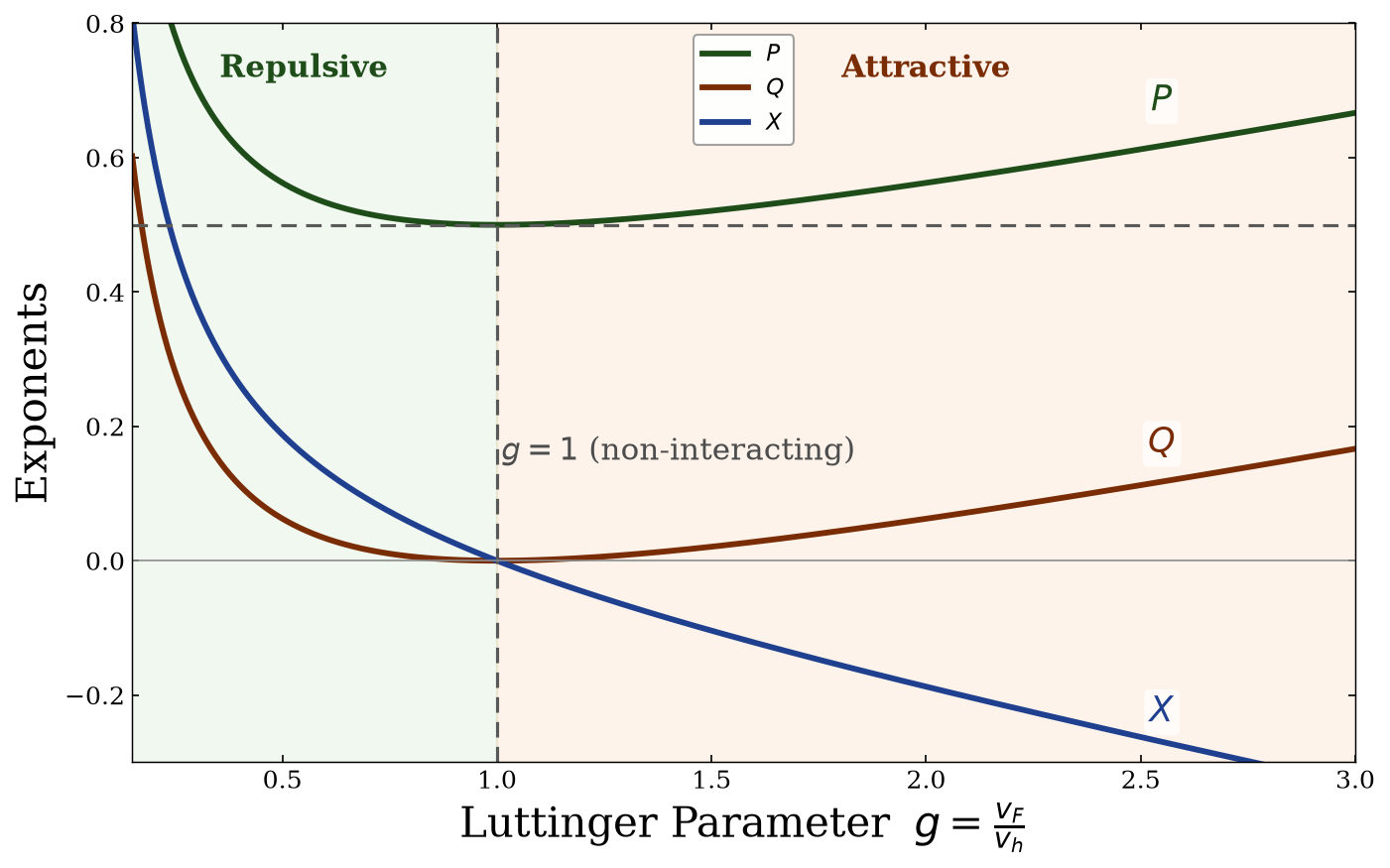}
\end{center}
\caption{Plot of correlation function exponents as a function of interaction dependent luttinger liquid parameter $g=v_F/v_h$.}
\label{PQX}
\end{figure}
Figure~(\ref{PQX}) shows the dependence of the exponents $P$, $Q$ and $X$ on the Luttinger parameter $g$. The vertical dotted line at $g=1$ separates the repulsive-interaction regime ($g<1$) from the attractive-interaction regime ($g>1$) and identifies the non-interacting limit. In the absence of interactions, $v_h$ becomes equal to $v_F$, and the exponent $P$ approaches $1/2$, as indicated by the horizontal dotted line, while $Q$ and $X$ vanish. As seen from the Green's function in equation~(\ref{GFall}), the holon contribution is governed by the exponent $P$ and propagates with the holon velocity $v_h$, whereas the corresponding spinon contribution carries an exponent $1/2$ and propagates with the Fermi velocity $v_F$. At $g=1$, the holon velocity becomes identical to the Fermi velocity, while $P=1/2$. Consequently, the holon and spinon factors become identical and combine to produce a single factor with unit exponent, thereby recovering the expected power-law behavior of the non-interacting two-point Green's function.

\subsection{Limiting Cases}
\noindent {\bf No interaction: }As a first consistency check, consider the non-interacting limit, for which the interaction strength is set to zero ($v_0 = 0$). In this case, the holon and spinon velocities become identical, $v_h=v_F$, and the exponents given in equation (\ref{exponents}) reduce to $P=1/2,\mbox{ } Q=0, \mbox{ }X=0.$
Substituting these values into equations (\ref{GFall})
\footnotesize
\begin{equation*}
\begin{aligned}
\Big\langle T\psi&_{R}(X_1)\psi_{R}^{\dagger}(X_2)\Big\rangle \sim  \sum_{\substack{\gamma,\gamma'\\=\pm1}}
\frac{ \delta_{\sigma,\sigma'}\,\theta(\gamma x_1)\theta(\gamma' x_2) \, g_{\gamma,\gamma'}(1,1)}{(x_1-x_2 -v_F \tau_{12})} \\
\end{aligned}
\end{equation*}
\normalsize
The remaining propagators, viz.,  $\langle T\psi_{L}(X_1)\psi_{L}^{\dagger}(X_2)\rangle$,  $\langle T\psi_{R}(X_1)\psi_{L}^{\dagger}(X_2)\rangle$ and  $\langle T\psi_{L}(X_1)\psi_{R}^{\dagger}(X_2)\rangle$ likewise reduce to their corresponding free-fermion expressions as given in equation (\ref{greenfunction0}). Therefore, the Green's functions obtained within UFBT correctly recover the non-interacting limit, while allowing for an arbitrary strength of the localized impurity cluster.\\

\noindent {\bf No impurity: }
It is important to note that the vanishing-barrier limit does not constitute a physically admissible limiting case of the present formulation. The Green's functions derived above correspond specifically to the sub-barrier regime, in which the particle energy satisfies $E<V_0$ (where $V_0$ is the barrier height) and transport across the localized barrier occurs through tunneling. If the barrier height is continuously reduced while the particle energy is kept fixed, the condition $E<V_0$ eventually ceases to hold and the system enters the above-barrier scattering regime, $E>V_0$. The latter represents a different physical regime and requires a corresponding scattering-state construction rather than the tunneling solution employed here. Consequently, the limit $V_0\rightarrow0$ cannot be used to assess the validity of the present tunneling Green's functions, since it lies outside the domain of applicability of the underlying formulation.\\

\noindent {\bf Half line: }A useful consistency check is obtained by considering the strong-barrier limit, $( |R|\rightarrow -1 )$ and $(T\rightarrow0)$, in which the localized barrier becomes perfectly reflecting and the system approaches the open-boundary limit. The corresponding boundary Green's function was obtained by Eggert et al. for an interacting one-dimensional electron system with an open boundary~\cite{eggert1996boundary}. In their notation, the full Green's function is expressed in terms of chiral Green's functions in equation (7) of their manuscript, with the latter given by equation (8) of their manuscript. Accordingly, the (RR) contribution is
\begin{equation}
\begin{aligned}
G_{RR}&(x,y,t)\propto\prod_{\nu=c,s}
\left[
\frac{4xy}
{v_\nu^2t^2-(x+y)^2}
\right]^{\frac{1}{8}
\left(\frac{1}{K_\nu^2}-K_\nu^2\right)}\\
&
\left(v_\nu t+x-y\right)^{-\frac{1}{8}
\left(K_\nu+\frac{1}{K_\nu}\right)^2}
\left(v_\nu t-x+y\right)^{-\frac{1}{8}
\left(K_\nu-\frac{1}{K_\nu}\right)^2}
\label{EggertGF}
\end{aligned}
\end{equation}
where (x) and (y) denote the distances of the two observation points from the boundary, \(v_\nu\) is the velocity in the charge \((\nu=c)\) or spin \((\nu=s)\) sector, and \(K_\nu\) is the corresponding Luttinger parameter. 
The correspondence with the UFBT Green's functions in equation (\ref{GFall}) becomes explicit by identifying the charge-sector Luttinger parameter as $K_c=v_F/v_h.$ Since the spinon velocity is simply the Fermi velocity, $v_s=v_F$, the corresponding spin-sector Luttinger parameter is $K_s=1$.
The exponents appearing in the UFBT expressions then become \footnotesize
\begin{equation}
P=\frac{1}{8}\left(K_c+\frac{1}{K_c} \right)^2
Q=\frac{1}{8}\left(K_c-\frac{1}{K_c} \right)^2
X=\frac{1}{8}\left(K_c^2-\frac{1}{K_c^2} \right)^2
\end{equation}
\normalsize
Hence, the charge-sector contribution to the UFBT right-right Green's function contains
\begin{equation}
\frac{(4x_1x_2)^{\frac{1}{8}\left(K_c^2-\frac{1}{K_c^2} \right)^2}}
{\large \substack{(x_1-x_2-v_h\tau_{12})^{\frac{1}{8}\left(K_c+\frac{1}{K_c} \right)^2}
(-x_1+x_2-v_h\tau_{12})^{\frac{1}{8}\left(K_c-\frac{1}{K_c} \right)^2}\\
(x_1+x_2-v_h\tau_{12})^{\frac{1}{8}\left(K_c^2-\frac{1}{K_c^2} \right)^2}
(-x_1-x_2-v_h\tau_{12})^{\frac{1}{8}\left(K_c^2-\frac{1}{K_c^2} \right)^2}}},
\end{equation}
\normalsize
while the spin sector contributes the factor
\begin{equation}
\frac{1}{(x_1-x_2-v_F\tau_{12})^{1/2}}.
\end{equation}
The charge- and spin-sector contributions given above, as obtained from the UFBT formalism, contain precisely the factors appearing in equation (\ref{EggertGF}), with $K_{\nu}=K_c=v_F/v_h$ for the charge (holon) sector and $K_{\nu}=K_s=1$ for the spin (spinon) sector. Thus, in the strong-barrier limit, the UFBT Green's function recovers the characteristic structure of the established open-boundary Green's function for an interacting one-dimensional electron system~\cite{eggert1996boundary}. This correspondence, including the consistent identification of the charge-sector parameter and the reduction of the spin sector to the expected free-spin contribution, provides a nontrivial consistency check of the UFBT Green's functions in the perfectly reflecting limit.\\

\noindent {\bf Far away from impurity: }
Consider the limit $x_1,x_2\gg 0$, with $x_1-x_2$ kept finite. For the $RR$ Green's function, the numerator factor $(4x_1x_2)$ and the two denominator factors involving $(x_1+x_2)$ have the same leading asymptotic dependence and therefore cancel, i.e.,
\begin{equation}
\frac{(4x_1x_2)^X}
{(x_1+x_2-v_h\tau_{12})^X(-x_1-x_2-v_h\tau_{12})^X}
\longrightarrow 1 ,
\end{equation}
so that
\begin{equation}
G_{RR}\propto
(x_1-x_2-v_h\tau_{12})^{-P}
(-x_1+x_2-v_h\tau_{12})^{-Q},
\end{equation}
which depends only on $x_1-x_2$ and $\tau_{12}$ and is therefore translationally invariant. The same argument applies to $G_{LL}$. In contrast, the mixed-chirality components \(G_{RL}\) and \(G_{LR}\) retain their \((x_1+x_2)\)-dependent impurity terms even in the far-field limit, since these are reflectional contributions that necessarily involve the region containing the impurity.

\subsection{Perturbative Comparison}
\noindent As a consistency check, a standard perturbation theory calculation is carried out in terms of the interaction strength $v_0$. Retaining terms up to first order in $v_0$, the Green's functions are obtained as (ss is same side of the impurity and os is opposite sides of impurity)
\footnotesize
\begin{equation*}
\begin{aligned}
G&_{RRss}(x_1,x_2,t_1,t_2)\\
=& \frac{i}{2\pi}\frac{1}{(x_1-x_2)-v_F(t_1-t_2)}+ \frac{i (t_1-t_2)}{4 \pi^2 ((x_1-x_2)-v_F(t_1-t_2))^2}v_0
\end{aligned}
\end{equation*}

\begin{equation*}
\begin{aligned}
G&_{RLss}(x_1,x_2,t_1,t_2)\\
 =& \frac{i{\bf R}}{2\pi}\frac{1}{(x_1+x_2)-v_F(t_1-t_2)}+  \frac{ i {\bf R} (t_1-t_2)}{4\pi^2((x_1+x_2)-v_F(t_1-t_2))^2}v_0
\end{aligned}
\end{equation*}

\begin{equation*}
\begin{aligned}
G&_{RRos}(x_1,x_2,t_1,t_2)\\
=& \frac{i{\bf T}}{2\pi}\frac{1}{(x_1-x_2)-v_F(t_1-t_2)}+ \frac{i {\bf T}(t_1-t_2)}{4 \pi^2 ((x_1-x_2)-v_F(t_1-t_2))^2}v_0
\end{aligned}
\end{equation*}
\normalsize
On the other hand, expanding the Green's functions obtained using UFBT in equation (\ref{GFall}) in powers of the interaction strength $v_0$ and retaining terms up to first order yields exactly the same expressions as those given above. Hence, the UFBT Green's functions are perturbatively consistent with the results obtained from standard perturbation theory.

\subsection{Validation through the Schwinger-Dyson Equation}

The Schwinger--Dyson equation provides a non-perturbative check on the Green's functions obtained within the UFBT formalism. It relates the full two-point Green's function for arbitrary chiralities $\nu$ and $\nu'$ to a four-point correlation function involving the interacting density. For the short-range interaction $v(x-y)=v_0\delta(x-y)$, the corresponding equations of motion are given by

\footnotesize
\begin{equation}
\begin{aligned}
(i\partial_t + i \nu &v_F   \partial_x) G^{full}_{\nu,\nu'}(x,x';t-t')\\
&=   \int dy \mbox{  }v(x-y)\mbox{  }   <T\mbox{  } \rho(y,t ) \psi_{\nu}(x,\sigma ,t)\psi^{\dagger}_{\nu'}(x',\sigma,t') >_{full}
\\
(-i\partial_{t'} -&i \nu'   v_F   \partial_{x'}) G^{full}_{\nu,\nu'}(x,x';t-t')\\
&=   \int dy \mbox{  }v(x'-y) \mbox{  }  <T\rho(y,t' ) \psi_{\nu}(x,\sigma ,t)\psi^{\dagger}_{\nu'}(x',\sigma,t') >_{full}
 \label{dyson2}
\end{aligned}
\end{equation}
\normalsize
Substitution of the UFBT Green's functions and the corresponding four-point correlation functions into equation (\ref{dyson2}) shows that the Schwinger--Dyson equations are satisfied identically for all combinations of $\nu$ and $\nu'$. The same-chirality and mixed-chirality components, including the reflection and transmission contributions, consistently satisfy the corresponding equations of motion. Thus, the UFBT Green's functions satisfy the Schwinger--Dyson equations, providing an independent non-perturbative validation of the UFBT formalism.

\subsection{N-point correlation functions}
The prescription given in equation (\ref{ucbt}) can be extended to calculate general $N$-point correlation functions. Consider the $N$-point function containing $N/2$ fermionic annihilation operators and $N/2$ fermionic creation operators,
\footnotesize
\begin{equation}
\begin{aligned}
\mathcal{G}^{(N)}
={}&
\left\langle
\psi_{\nu_1}(X_1)
\cdots
\psi_{\nu_{\frac{N}{2}}}(X_{\frac{N}{2}})
\psi_{\nu_{\frac{N}{2}+1}}^{\dagger}(X_{\frac{N}{2}+1})
\cdots
\psi_{\nu_N}^{\dagger}(X_N)
\right\rangle 
\end{aligned}
\label{Npoint}
\end{equation}
\normalsize
where $X_i=(x_i,\sigma_i,t_i)$ denotes the space, spin, and time coordinates of the $i$-th operator. Using the prescription in equation (\ref{ucbt}), each fermionic operator is represented by an exponential of the corresponding bosonic field. For convenience, define
\begin{equation}
B_i =
\begin{cases}
\mbox{ }\mbox{ }i\Theta_{\nu_i}(X_i),&  \hspace{1cm}1\leq i\leq \frac{N}{2},\\[4pt]
-i\Theta_{\nu_i}(X_i), &  \hspace{1cm}\frac{N}{2}<i\leq N.
\end{cases}
\label{Bi}
\end{equation}
where  $\Theta_{\nu}(X_i) =(1/\sqrt{2}) ( \theta_\nu(x_i,\sigma_i,t_i)+\theta_\nu(-x_i,\sigma_i,t_i))$ as already defined in an earlier section. The $N$-point correlation function can then be written as
\begin{equation} \hspace{2cm}
\mathcal{G}^{(N)}
\sim
\left\langle
\prod_{i=1}^{N} e^{B_i}
\right\rangle .
\label{Npoint_exp}
\end{equation}
For Gaussian bosonic fields, the Baker--Campbell--Hausdorff (BCH) relation gives
\begin{equation}
\mathcal{G}^{(N)}
\sim\exp
\left[
\frac{1}{2}
\sum_{i=1}^{N}
\left\langle B_i^2 \right\rangle
+
\sum_{i<j}
\left\langle B_i B_j \right\rangle
\right].
\label{Npoint_BCH}
\end{equation}
Equivalently, the above expression can be written in product form as
\begin{equation}
\mathcal{G}^{(N)}
\sim\prod_{i=1}^{N}
\exp\left[
\frac{1}{2}
\left\langle B_i^2 \right\rangle
\right]
\prod_{i<j}
\exp\left[
\left\langle B_i B_j \right\rangle
\right].
\label{Npoint_product}
\end{equation}
Thus, the general $N$-point correlation function is determined by the self-contractions and all pairwise contractions of the bosonic fields. The two-point Green's function considered earlier is recovered as the special case $N=2$.

\subsection{Applications}
The correlation functions obtained within the present formalism provide a framework for investigating a range of physical properties of inhomogeneous one-dimensional quantum systems. Since the correlation function exponents are independent of the impurity strength, the resulting conductance exhibits universal interaction-dependent scaling for a given Luttinger parameter $K$, consistent with the Kane--Fisher picture of impurity effects in Luttinger liquids \cite{kane1992transport} and with the boundary conformal field theory description \cite{eggert1996boundary}. The two-point Green's functions can therefore be used to study transport phenomena such as conductance and resonant tunneling, as well as spectroscopic quantities including the local or dynamical density of states. The complete spatial dependence of the Green's functions also permits the calculation of Friedel oscillations and other spatial correlation effects, while the four-point correlation functions provide access to higher-order density and current current correlations. In particular, the formalism can be applied to systems exhibiting the crossover between the `cutting'' and `healing'' of a quantum chain associated with weak and strong impurity limits \cite{kane1992transport}. Beyond a single localized impurity, the same prescription can be applied to more general inhomogeneous configurations, including fermionic ladder systems and systems containing slowly moving impurities. The resulting analytical correlation functions can further serve as benchmarks for numerical approaches such as density-matrix renormalization group (DMRG) calculations \cite{white1992density,schollwock2005density}, particularly in regimes where strong interactions and spatial inhomogeneity make analytical treatments difficult.
\section{Conclusions}
The Unified Field Bosonization Technique (UFBT) provides a unified analytical framework for obtaining closed-form expressions for the leading singular contributions to Green's functions of strongly interacting one-dimensional fermionic systems in the presence of arbitrary localized static impurity configurations. By employing a symmetrized bosonic-field prescription that explicitly incorporates the impurity position, UFBT captures the breaking of translational invariance while retaining the non-perturbative structure of the Luttinger-liquid description without requiring renormalization-group resummation. The formalism yields Green's functions for localized impurity clusters containing combinations of barriers and wells, with the impurity dependence entering through the corresponding reflection and transmission amplitudes. The resulting correlation functions retain the characteristic power-law behavior of a Luttinger liquid, with exponents determined solely by the bulk Luttinger parameter and independent of the impurity strength, while the impurity modifies the spatial structure and amplitudes of the correlations. The consistency of the formalism is established through several complementary checks, including known limiting cases, comparison with perturbative results, and verification through the Schwinger-Dyson equation. These results demonstrate that UFBT provides a consistent and non-perturbative description of strongly inhomogeneous Luttinger liquids with localized impurities, while offering closed-form analytical benchmarks for further studies of interacting one-dimensional systems. The framework can also be extended to more complex impurity configurations and other inhomogeneous Luttinger-liquid settings.\\

\section*{Acknowledgments}
The author gratefully acknowledges Prof. Girish S. Setlur for his guidance and valuable discussions during the development of the ideas underlying this work.

\bibliographystyle{apsrev4-1}
\bibliography{ref}

\begin{thebibliography}{39}%
\makeatletter
\providecommand \@ifxundefined [1]{%
 \@ifx{#1\undefined}
}%
\providecommand \@ifnum [1]{%
 \ifnum #1\expandafter \@firstoftwo
 \else \expandafter \@secondoftwo
 \fi
}%
\providecommand \@ifx [1]{%
 \ifx #1\expandafter \@firstoftwo
 \else \expandafter \@secondoftwo
 \fi
}%
\providecommand \natexlab [1]{#1}%
\providecommand \enquote  [1]{``#1''}%
\providecommand \bibnamefont  [1]{#1}%
\providecommand \bibfnamefont [1]{#1}%
\providecommand \citenamefont [1]{#1}%
\providecommand \href@noop [0]{\@secondoftwo}%
\providecommand \href [0]{\begingroup \@sanitize@url \@href}%
\providecommand \@href[1]{\@@startlink{#1}\@@href}%
\providecommand \@@href[1]{\endgroup#1\@@endlink}%
\providecommand \@sanitize@url [0]{\catcode `\\12\catcode `\$12\catcode
  `\&12\catcode `\#12\catcode `\^12\catcode `\_12\catcode `\%12\relax}%
\providecommand \@@startlink[1]{}%
\providecommand \@@endlink[0]{}%
\providecommand \url  [0]{\begingroup\@sanitize@url \@url }%
\providecommand \@url [1]{\endgroup\@href {#1}{\urlprefix }}%
\providecommand \urlprefix  [0]{URL }%
\providecommand \Eprint [0]{\href }%
\providecommand \doibase [0]{http://dx.doi.org/}%
\providecommand \selectlanguage [0]{\@gobble}%
\providecommand \bibinfo  [0]{\@secondoftwo}%
\providecommand \bibfield  [0]{\@secondoftwo}%
\providecommand \translation [1]{[#1]}%
\providecommand \BibitemOpen [0]{}%
\providecommand \bibitemStop [0]{}%
\providecommand \bibitemNoStop [0]{.\EOS\space}%
\providecommand \EOS [0]{\spacefactor3000\relax}%
\providecommand \BibitemShut  [1]{\csname bibitem#1\endcsname}%
\let\auto@bib@innerbib\@empty
\bibitem [{\citenamefont {Bahovadinov}\ and\ \citenamefont
  {Matveenko}(2022)}]{bahovadinov2022effects}%
  \BibitemOpen
  \bibfield  {author} {\bibinfo {author} {\bibfnamefont {M.}~\bibnamefont
  {Bahovadinov}}\ and\ \bibinfo {author} {\bibfnamefont {S.}~\bibnamefont
  {Matveenko}},\ }\href@noop {} {\bibfield  {journal} {\bibinfo  {journal}
  {Journal of Physics: Condensed Matter}\ }\textbf {\bibinfo {volume} {34}},\
  \bibinfo {pages} {315601} (\bibinfo {year} {2022})}\BibitemShut {NoStop}%
\bibitem [{\citenamefont {Gluza}\ \emph {et~al.}(2022)\citenamefont {Gluza},
  \citenamefont {Moosavi},\ and\ \citenamefont
  {Sotiriadis}}]{gluza2022breaking}%
  \BibitemOpen
  \bibfield  {author} {\bibinfo {author} {\bibfnamefont {M.}~\bibnamefont
  {Gluza}}, \bibinfo {author} {\bibfnamefont {P.}~\bibnamefont {Moosavi}}, \
  and\ \bibinfo {author} {\bibfnamefont {S.}~\bibnamefont {Sotiriadis}},\
  }\href@noop {} {\bibfield  {journal} {\bibinfo  {journal} {Journal of Physics
  A: Mathematical and Theoretical}\ }\textbf {\bibinfo {volume} {55}},\
  \bibinfo {pages} {054002} (\bibinfo {year} {2022})}\BibitemShut {NoStop}%
\bibitem [{\citenamefont {Gotta}\ and\ \citenamefont
  {Giamarchi}(2024)}]{gotta2024dirac}%
  \BibitemOpen
  \bibfield  {author} {\bibinfo {author} {\bibfnamefont {L.}~\bibnamefont
  {Gotta}}\ and\ \bibinfo {author} {\bibfnamefont {T.}~\bibnamefont
  {Giamarchi}},\ }\href@noop {} {\bibfield  {journal} {\bibinfo  {journal}
  {Physical Review B}\ }\textbf {\bibinfo {volume} {110}},\ \bibinfo {pages}
  {165116} (\bibinfo {year} {2024})}\BibitemShut {NoStop}%
\bibitem [{\citenamefont {Giamarchi}(2004)}]{giamarchi2004quantum}%
  \BibitemOpen
  \bibfield  {author} {\bibinfo {author} {\bibfnamefont {T.}~\bibnamefont
  {Giamarchi}},\ }\href@noop {} {\emph {\bibinfo {title} {Quantum physics in
  one dimension}}},\ Vol.\ \bibinfo {volume} {121}\ (\bibinfo  {publisher}
  {Clarendon Oxford},\ \bibinfo {year} {2004})\BibitemShut {NoStop}%
\bibitem [{\citenamefont {Furusaki}(2005)}]{furusaki2005kondo}%
  \BibitemOpen
  \bibfield  {author} {\bibinfo {author} {\bibfnamefont {A.}~\bibnamefont
  {Furusaki}},\ }\href@noop {} {\bibfield  {journal} {\bibinfo  {journal}
  {Journal of the Physical Society of Japan}\ }\textbf {\bibinfo {volume}
  {74}},\ \bibinfo {pages} {73} (\bibinfo {year} {2005})}\BibitemShut {NoStop}%
\bibitem [{\citenamefont {Das}\ and\ \citenamefont
  {Setlur}(2019{\natexlab{a}})}]{das2019transport}%
  \BibitemOpen
  \bibfield  {author} {\bibinfo {author} {\bibfnamefont {J.~P.}\ \bibnamefont
  {Das}}\ and\ \bibinfo {author} {\bibfnamefont {G.~S.}\ \bibnamefont
  {Setlur}},\ }\href@noop {} {\bibfield  {journal} {\bibinfo  {journal}
  {Physica E: Low-dimensional Systems and Nanostructures}\ }\textbf {\bibinfo
  {volume} {110}},\ \bibinfo {pages} {39} (\bibinfo {year}
  {2019}{\natexlab{a}})}\BibitemShut {NoStop}%
\bibitem [{\citenamefont {Kane}\ and\ \citenamefont
  {Fisher}(1992)}]{kane1992transport}%
  \BibitemOpen
  \bibfield  {author} {\bibinfo {author} {\bibfnamefont {C.}~\bibnamefont
  {Kane}}\ and\ \bibinfo {author} {\bibfnamefont {M.~P.}\ \bibnamefont
  {Fisher}},\ }\href@noop {} {\bibfield  {journal} {\bibinfo  {journal}
  {Physical Review Letters}\ }\textbf {\bibinfo {volume} {68}},\ \bibinfo
  {pages} {1220} (\bibinfo {year} {1992})}\BibitemShut {NoStop}%
\bibitem [{\citenamefont {Artemenko}\ and\ \citenamefont
  {Remizov}(2005)}]{artemenko2005low}%
  \BibitemOpen
  \bibfield  {author} {\bibinfo {author} {\bibfnamefont {S.~N.}\ \bibnamefont
  {Artemenko}}\ and\ \bibinfo {author} {\bibfnamefont {S.}~\bibnamefont
  {Remizov}},\ }\href@noop {} {\bibfield  {journal} {\bibinfo  {journal}
  {Physical Review B}\ }\textbf {\bibinfo {volume} {72}},\ \bibinfo {pages}
  {125118} (\bibinfo {year} {2005})}\BibitemShut {NoStop}%
\bibitem [{\citenamefont {Dinh}\ \emph {et~al.}(2010)\citenamefont {Dinh},
  \citenamefont {Bagrets},\ and\ \citenamefont {Mirlin}}]{dinh2010tunneling}%
  \BibitemOpen
  \bibfield  {author} {\bibinfo {author} {\bibfnamefont {S.~N.}\ \bibnamefont
  {Dinh}}, \bibinfo {author} {\bibfnamefont {D.~A.}\ \bibnamefont {Bagrets}}, \
  and\ \bibinfo {author} {\bibfnamefont {A.~D.}\ \bibnamefont {Mirlin}},\
  }\href@noop {} {\bibfield  {journal} {\bibinfo  {journal} {Physical Review
  B}\ }\textbf {\bibinfo {volume} {81}},\ \bibinfo {pages} {081306} (\bibinfo
  {year} {2010})}\BibitemShut {NoStop}%
\bibitem [{\citenamefont {Kainaris}\ \emph {et~al.}(2018)\citenamefont
  {Kainaris}, \citenamefont {Carr},\ and\ \citenamefont
  {Mirlin}}]{kainaris2018transmission}%
  \BibitemOpen
  \bibfield  {author} {\bibinfo {author} {\bibfnamefont {N.}~\bibnamefont
  {Kainaris}}, \bibinfo {author} {\bibfnamefont {S.~T.}\ \bibnamefont {Carr}},
  \ and\ \bibinfo {author} {\bibfnamefont {A.~D.}\ \bibnamefont {Mirlin}},\
  }\href@noop {} {\bibfield  {journal} {\bibinfo  {journal} {Physical Review
  B}\ }\textbf {\bibinfo {volume} {97}},\ \bibinfo {pages} {115107} (\bibinfo
  {year} {2018})}\BibitemShut {NoStop}%
\bibitem [{\citenamefont {Lo}\ \emph {et~al.}(2019)\citenamefont {Lo},
  \citenamefont {Fukusumi}, \citenamefont {Oshikawa}, \citenamefont {Kao},
  \citenamefont {Chen} \emph {et~al.}}]{lo2019crossover}%
  \BibitemOpen
  \bibfield  {author} {\bibinfo {author} {\bibfnamefont {C.-Y.}\ \bibnamefont
  {Lo}}, \bibinfo {author} {\bibfnamefont {Y.}~\bibnamefont {Fukusumi}},
  \bibinfo {author} {\bibfnamefont {M.}~\bibnamefont {Oshikawa}}, \bibinfo
  {author} {\bibfnamefont {Y.-J.}\ \bibnamefont {Kao}}, \bibinfo {author}
  {\bibfnamefont {P.}~\bibnamefont {Chen}},  \emph {et~al.},\ }\href@noop {}
  {\bibfield  {journal} {\bibinfo  {journal} {Physical Review B}\ }\textbf
  {\bibinfo {volume} {99}},\ \bibinfo {pages} {121103} (\bibinfo {year}
  {2019})}\BibitemShut {NoStop}%
\bibitem [{\citenamefont {Grishin}\ \emph {et~al.}(2004)\citenamefont
  {Grishin}, \citenamefont {Yurkevich},\ and\ \citenamefont
  {Lerner}}]{grishin2004functional}%
  \BibitemOpen
  \bibfield  {author} {\bibinfo {author} {\bibfnamefont {A.}~\bibnamefont
  {Grishin}}, \bibinfo {author} {\bibfnamefont {I.~V.}\ \bibnamefont
  {Yurkevich}}, \ and\ \bibinfo {author} {\bibfnamefont {I.~V.}\ \bibnamefont
  {Lerner}},\ }\href@noop {} {\bibfield  {journal} {\bibinfo  {journal}
  {Physical Review B}\ }\textbf {\bibinfo {volume} {69}},\ \bibinfo {pages}
  {165108} (\bibinfo {year} {2004})}\BibitemShut {NoStop}%
\bibitem [{\citenamefont {Matveev}\ \emph {et~al.}(1993)\citenamefont
  {Matveev}, \citenamefont {Yue},\ and\ \citenamefont
  {Glazman}}]{matveev1993tunneling}%
  \BibitemOpen
  \bibfield  {author} {\bibinfo {author} {\bibfnamefont {K.}~\bibnamefont
  {Matveev}}, \bibinfo {author} {\bibfnamefont {D.}~\bibnamefont {Yue}}, \ and\
  \bibinfo {author} {\bibfnamefont {L.}~\bibnamefont {Glazman}},\ }\href@noop
  {} {\bibfield  {journal} {\bibinfo  {journal} {Physical review letters}\
  }\textbf {\bibinfo {volume} {71}},\ \bibinfo {pages} {3351} (\bibinfo {year}
  {1993})}\BibitemShut {NoStop}%
\bibitem [{\citenamefont {Das}\ and\ \citenamefont
  {Setlur}(2018{\natexlab{a}})}]{das2018ponderous}%
  \BibitemOpen
  \bibfield  {author} {\bibinfo {author} {\bibfnamefont {J.~P.}\ \bibnamefont
  {Das}}\ and\ \bibinfo {author} {\bibfnamefont {G.~S.}\ \bibnamefont
  {Setlur}},\ }\href@noop {} {\bibfield  {journal} {\bibinfo  {journal} {EPL
  (Europhysics Letters)}\ }\textbf {\bibinfo {volume} {123}},\ \bibinfo {pages}
  {27002} (\bibinfo {year} {2018}{\natexlab{a}})}\BibitemShut {NoStop}%
\bibitem [{\citenamefont {Qin}\ \emph {et~al.}(1996)\citenamefont {Qin},
  \citenamefont {Fabrizio},\ and\ \citenamefont {Yu}}]{qin1996impurity}%
  \BibitemOpen
  \bibfield  {author} {\bibinfo {author} {\bibfnamefont {S.}~\bibnamefont
  {Qin}}, \bibinfo {author} {\bibfnamefont {M.}~\bibnamefont {Fabrizio}}, \
  and\ \bibinfo {author} {\bibfnamefont {L.}~\bibnamefont {Yu}},\ }\href@noop
  {} {\bibfield  {journal} {\bibinfo  {journal} {Physical Review B}\ }\textbf
  {\bibinfo {volume} {54}},\ \bibinfo {pages} {R9643} (\bibinfo {year}
  {1996})}\BibitemShut {NoStop}%
\bibitem [{\citenamefont {Hamamoto}\ \emph {et~al.}(2008)\citenamefont
  {Hamamoto}, \citenamefont {Imura},\ and\ \citenamefont
  {Kato}}]{hamamoto2008numerical}%
  \BibitemOpen
  \bibfield  {author} {\bibinfo {author} {\bibfnamefont {Y.}~\bibnamefont
  {Hamamoto}}, \bibinfo {author} {\bibfnamefont {K.-I.}\ \bibnamefont {Imura}},
  \ and\ \bibinfo {author} {\bibfnamefont {T.}~\bibnamefont {Kato}},\
  }\href@noop {} {\bibfield  {journal} {\bibinfo  {journal} {Physical Review
  B}\ }\textbf {\bibinfo {volume} {77}},\ \bibinfo {pages} {165402} (\bibinfo
  {year} {2008})}\BibitemShut {NoStop}%
\bibitem [{\citenamefont {Freyn}\ and\ \citenamefont
  {Florens}(2011)}]{freyn2011numerical}%
  \BibitemOpen
  \bibfield  {author} {\bibinfo {author} {\bibfnamefont {A.}~\bibnamefont
  {Freyn}}\ and\ \bibinfo {author} {\bibfnamefont {S.}~\bibnamefont
  {Florens}},\ }\href@noop {} {\bibfield  {journal} {\bibinfo  {journal}
  {Physical Review Letters}\ }\textbf {\bibinfo {volume} {107}},\ \bibinfo
  {pages} {017201} (\bibinfo {year} {2011})}\BibitemShut {NoStop}%
\bibitem [{\citenamefont {Moon}\ \emph {et~al.}(1993)\citenamefont {Moon},
  \citenamefont {Yi}, \citenamefont {Kane}, \citenamefont {Girvin},\ and\
  \citenamefont {Fisher}}]{moon1993resonant}%
  \BibitemOpen
  \bibfield  {author} {\bibinfo {author} {\bibfnamefont {K.}~\bibnamefont
  {Moon}}, \bibinfo {author} {\bibfnamefont {H.}~\bibnamefont {Yi}}, \bibinfo
  {author} {\bibfnamefont {C.}~\bibnamefont {Kane}}, \bibinfo {author}
  {\bibfnamefont {S.}~\bibnamefont {Girvin}}, \ and\ \bibinfo {author}
  {\bibfnamefont {M.~P.}\ \bibnamefont {Fisher}},\ }\href@noop {} {\bibfield
  {journal} {\bibinfo  {journal} {Physical review letters}\ }\textbf {\bibinfo
  {volume} {71}},\ \bibinfo {pages} {4381} (\bibinfo {year}
  {1993})}\BibitemShut {NoStop}%
\bibitem [{\citenamefont {Ishii}\ \emph {et~al.}(2003)\citenamefont {Ishii},
  \citenamefont {Kataura}, \citenamefont {Shiozawa}, \citenamefont {Yoshioka},
  \citenamefont {Otsubo}, \citenamefont {Takayama}, \citenamefont {Miyahara},
  \citenamefont {Suzuki}, \citenamefont {Achiba}, \citenamefont {Nakatake}
  \emph {et~al.}}]{ishii2003direct}%
  \BibitemOpen
  \bibfield  {author} {\bibinfo {author} {\bibfnamefont {H.}~\bibnamefont
  {Ishii}}, \bibinfo {author} {\bibfnamefont {H.}~\bibnamefont {Kataura}},
  \bibinfo {author} {\bibfnamefont {H.}~\bibnamefont {Shiozawa}}, \bibinfo
  {author} {\bibfnamefont {H.}~\bibnamefont {Yoshioka}}, \bibinfo {author}
  {\bibfnamefont {H.}~\bibnamefont {Otsubo}}, \bibinfo {author} {\bibfnamefont
  {Y.}~\bibnamefont {Takayama}}, \bibinfo {author} {\bibfnamefont
  {T.}~\bibnamefont {Miyahara}}, \bibinfo {author} {\bibfnamefont
  {S.}~\bibnamefont {Suzuki}}, \bibinfo {author} {\bibfnamefont
  {Y.}~\bibnamefont {Achiba}}, \bibinfo {author} {\bibfnamefont
  {M.}~\bibnamefont {Nakatake}},  \emph {et~al.},\ }\href@noop {} {\bibfield
  {journal} {\bibinfo  {journal} {Nature}\ }\textbf {\bibinfo {volume} {426}},\
  \bibinfo {pages} {540} (\bibinfo {year} {2003})}\BibitemShut {NoStop}%
\bibitem [{\citenamefont {Auslaender}\ \emph {et~al.}(2002)\citenamefont
  {Auslaender}, \citenamefont {Yacoby}, \citenamefont {De~Picciotto},
  \citenamefont {Baldwin}, \citenamefont {Pfeiffer},\ and\ \citenamefont
  {West}}]{auslaender2002tunneling}%
  \BibitemOpen
  \bibfield  {author} {\bibinfo {author} {\bibfnamefont {O.}~\bibnamefont
  {Auslaender}}, \bibinfo {author} {\bibfnamefont {A.}~\bibnamefont {Yacoby}},
  \bibinfo {author} {\bibfnamefont {R.}~\bibnamefont {De~Picciotto}}, \bibinfo
  {author} {\bibfnamefont {K.}~\bibnamefont {Baldwin}}, \bibinfo {author}
  {\bibfnamefont {L.}~\bibnamefont {Pfeiffer}}, \ and\ \bibinfo {author}
  {\bibfnamefont {K.}~\bibnamefont {West}},\ }\href@noop {} {\bibfield
  {journal} {\bibinfo  {journal} {Science}\ }\textbf {\bibinfo {volume}
  {295}},\ \bibinfo {pages} {825} (\bibinfo {year} {2002})}\BibitemShut
  {NoStop}%
\bibitem [{\citenamefont {Glazman}\ and\ \citenamefont
  {Larkin}(1997)}]{glazman1997new}%
  \BibitemOpen
  \bibfield  {author} {\bibinfo {author} {\bibfnamefont {L.}~\bibnamefont
  {Glazman}}\ and\ \bibinfo {author} {\bibfnamefont {A.}~\bibnamefont
  {Larkin}},\ }\href@noop {} {\bibfield  {journal} {\bibinfo  {journal}
  {Physical review letters}\ }\textbf {\bibinfo {volume} {79}},\ \bibinfo
  {pages} {3736} (\bibinfo {year} {1997})}\BibitemShut {NoStop}%
\bibitem [{\citenamefont {Bloch}\ \emph {et~al.}(2008)\citenamefont {Bloch},
  \citenamefont {Dalibard},\ and\ \citenamefont {Zwerger}}]{bloch2008many}%
  \BibitemOpen
  \bibfield  {author} {\bibinfo {author} {\bibfnamefont {I.}~\bibnamefont
  {Bloch}}, \bibinfo {author} {\bibfnamefont {J.}~\bibnamefont {Dalibard}}, \
  and\ \bibinfo {author} {\bibfnamefont {W.}~\bibnamefont {Zwerger}},\
  }\href@noop {} {\bibfield  {journal} {\bibinfo  {journal} {Reviews of modern
  physics}\ }\textbf {\bibinfo {volume} {80}},\ \bibinfo {pages} {885}
  (\bibinfo {year} {2008})}\BibitemShut {NoStop}%
\bibitem [{\citenamefont {Schwartz}\ \emph {et~al.}(1998)\citenamefont
  {Schwartz}, \citenamefont {Dressel}, \citenamefont {Gr{\"u}ner},
  \citenamefont {Vescoli}, \citenamefont {Degiorgi},\ and\ \citenamefont
  {Giamarchi}}]{schwartz1998chain}%
  \BibitemOpen
  \bibfield  {author} {\bibinfo {author} {\bibfnamefont {A.}~\bibnamefont
  {Schwartz}}, \bibinfo {author} {\bibfnamefont {M.}~\bibnamefont {Dressel}},
  \bibinfo {author} {\bibfnamefont {G.}~\bibnamefont {Gr{\"u}ner}}, \bibinfo
  {author} {\bibfnamefont {V.}~\bibnamefont {Vescoli}}, \bibinfo {author}
  {\bibfnamefont {L.}~\bibnamefont {Degiorgi}}, \ and\ \bibinfo {author}
  {\bibfnamefont {T.}~\bibnamefont {Giamarchi}},\ }\href@noop {} {\bibfield
  {journal} {\bibinfo  {journal} {Physical Review B}\ }\textbf {\bibinfo
  {volume} {58}},\ \bibinfo {pages} {1261} (\bibinfo {year}
  {1998})}\BibitemShut {NoStop}%
\bibitem [{\citenamefont {Dagotto}(1999)}]{dagotto1999experiments}%
  \BibitemOpen
  \bibfield  {author} {\bibinfo {author} {\bibfnamefont {E.}~\bibnamefont
  {Dagotto}},\ }\href@noop {} {\bibfield  {journal} {\bibinfo  {journal}
  {Reports on Progress in Physics}\ }\textbf {\bibinfo {volume} {62}},\
  \bibinfo {pages} {1525} (\bibinfo {year} {1999})}\BibitemShut {NoStop}%
\bibitem [{\citenamefont {Das}\ and\ \citenamefont
  {Setlur}(2018{\natexlab{b}})}]{das2018quantum}%
  \BibitemOpen
  \bibfield  {author} {\bibinfo {author} {\bibfnamefont {J.~P.}\ \bibnamefont
  {Das}}\ and\ \bibinfo {author} {\bibfnamefont {G.~S.}\ \bibnamefont
  {Setlur}},\ }\href@noop {} {\bibfield  {journal} {\bibinfo  {journal}
  {International Journal of Modern Physics A}\ ,\ \bibinfo {pages} {1850174}}
  (\bibinfo {year} {2018}{\natexlab{b}})}\BibitemShut {NoStop}%
\bibitem [{\citenamefont {Eggert}\ \emph {et~al.}(1996)\citenamefont {Eggert},
  \citenamefont {Johannesson},\ and\ \citenamefont
  {Mattsson}}]{eggert1996boundary}%
  \BibitemOpen
  \bibfield  {author} {\bibinfo {author} {\bibfnamefont {S.}~\bibnamefont
  {Eggert}}, \bibinfo {author} {\bibfnamefont {H.}~\bibnamefont {Johannesson}},
  \ and\ \bibinfo {author} {\bibfnamefont {A.}~\bibnamefont {Mattsson}},\
  }\href@noop {} {\bibfield  {journal} {\bibinfo  {journal} {Physical review
  letters}\ }\textbf {\bibinfo {volume} {76}},\ \bibinfo {pages} {1505}
  (\bibinfo {year} {1996})}\BibitemShut {NoStop}%
\bibitem [{\citenamefont {Haldane}(1981)}]{haldane1981luttinger}%
  \BibitemOpen
  \bibfield  {author} {\bibinfo {author} {\bibfnamefont {F.}~\bibnamefont
  {Haldane}},\ }\href@noop {} {\bibfield  {journal} {\bibinfo  {journal}
  {Journal of Physics C: Solid State Physics}\ }\textbf {\bibinfo {volume}
  {14}},\ \bibinfo {pages} {2585} (\bibinfo {year} {1981})}\BibitemShut
  {NoStop}%
\bibitem [{\citenamefont {Von~Delft}\ and\ \citenamefont
  {Schoeller}(1998)}]{von1998bosonization}%
  \BibitemOpen
  \bibfield  {author} {\bibinfo {author} {\bibfnamefont {J.}~\bibnamefont
  {Von~Delft}}\ and\ \bibinfo {author} {\bibfnamefont {H.}~\bibnamefont
  {Schoeller}},\ }\href@noop {} {\bibfield  {journal} {\bibinfo  {journal}
  {Annalen der Physik}\ }\textbf {\bibinfo {volume} {7}},\ \bibinfo {pages}
  {225} (\bibinfo {year} {1998})}\BibitemShut {NoStop}%
\bibitem [{\citenamefont {White}(1992)}]{white1992density}%
  \BibitemOpen
  \bibfield  {author} {\bibinfo {author} {\bibfnamefont {S.~R.}\ \bibnamefont
  {White}},\ }\href@noop {} {\bibfield  {journal} {\bibinfo  {journal}
  {Physical review letters}\ }\textbf {\bibinfo {volume} {69}},\ \bibinfo
  {pages} {2863} (\bibinfo {year} {1992})}\BibitemShut {NoStop}%
\bibitem [{\citenamefont {Schollw{\"o}ck}(2005)}]{schollwock2005density}%
  \BibitemOpen
  \bibfield  {author} {\bibinfo {author} {\bibfnamefont {U.}~\bibnamefont
  {Schollw{\"o}ck}},\ }\href@noop {} {\bibfield  {journal} {\bibinfo  {journal}
  {Reviews of modern physics}\ }\textbf {\bibinfo {volume} {77}},\ \bibinfo
  {pages} {259} (\bibinfo {year} {2005})}\BibitemShut {NoStop}%
\bibitem [{\citenamefont {Qin}\ \emph {et~al.}(1997)\citenamefont {Qin},
  \citenamefont {Fabrizio}, \citenamefont {Yu}, \citenamefont {Oshikawa},\ and\
  \citenamefont {Affleck}}]{qin1997impurity}%
  \BibitemOpen
  \bibfield  {author} {\bibinfo {author} {\bibfnamefont {S.}~\bibnamefont
  {Qin}}, \bibinfo {author} {\bibfnamefont {M.}~\bibnamefont {Fabrizio}},
  \bibinfo {author} {\bibfnamefont {L.}~\bibnamefont {Yu}}, \bibinfo {author}
  {\bibfnamefont {M.}~\bibnamefont {Oshikawa}}, \ and\ \bibinfo {author}
  {\bibfnamefont {I.}~\bibnamefont {Affleck}},\ }\href@noop {} {\bibfield
  {journal} {\bibinfo  {journal} {Physical Review B}\ }\textbf {\bibinfo
  {volume} {56}},\ \bibinfo {pages} {9766} (\bibinfo {year}
  {1997})}\BibitemShut {NoStop}%
\bibitem [{\citenamefont {Schollw{\"o}ck}\ \emph {et~al.}(2002)\citenamefont
  {Schollw{\"o}ck}, \citenamefont {Meden}, \citenamefont {Metzner},\ and\
  \citenamefont {Sch{\"o}nhammer}}]{schollwock2002dmrg}%
  \BibitemOpen
  \bibfield  {author} {\bibinfo {author} {\bibfnamefont {U.}~\bibnamefont
  {Schollw{\"o}ck}}, \bibinfo {author} {\bibfnamefont {V.}~\bibnamefont
  {Meden}}, \bibinfo {author} {\bibfnamefont {W.}~\bibnamefont {Metzner}}, \
  and\ \bibinfo {author} {\bibfnamefont {K.}~\bibnamefont {Sch{\"o}nhammer}},\
  }\href@noop {} {\bibfield  {journal} {\bibinfo  {journal} {Progress of
  Theoretical Physics Supplement}\ }\textbf {\bibinfo {volume} {145}},\
  \bibinfo {pages} {312} (\bibinfo {year} {2002})}\BibitemShut {NoStop}%
\bibitem [{\citenamefont {Bischoff}\ and\ \citenamefont
  {Jeckelmann}(2017)}]{bischoff2017density}%
  \BibitemOpen
  \bibfield  {author} {\bibinfo {author} {\bibfnamefont {J.-M.}\ \bibnamefont
  {Bischoff}}\ and\ \bibinfo {author} {\bibfnamefont {E.}~\bibnamefont
  {Jeckelmann}},\ }\href@noop {} {\bibfield  {journal} {\bibinfo  {journal}
  {Physical Review B}\ }\textbf {\bibinfo {volume} {96}},\ \bibinfo {pages}
  {195111} (\bibinfo {year} {2017})}\BibitemShut {NoStop}%
\bibitem [{\citenamefont {Voit}(1995)}]{voit1995one}%
  \BibitemOpen
  \bibfield  {author} {\bibinfo {author} {\bibfnamefont {J.}~\bibnamefont
  {Voit}},\ }\href@noop {} {\bibfield  {journal} {\bibinfo  {journal} {Reports
  on Progress in Physics}\ }\textbf {\bibinfo {volume} {58}},\ \bibinfo {pages}
  {977} (\bibinfo {year} {1995})}\BibitemShut {NoStop}%
\bibitem [{\citenamefont {Das}\ and\ \citenamefont
  {Setlur}(2019{\natexlab{b}})}]{das2019conductance}%
  \BibitemOpen
  \bibfield  {author} {\bibinfo {author} {\bibfnamefont {J.~P.}\ \bibnamefont
  {Das}}\ and\ \bibinfo {author} {\bibfnamefont {G.~S.}\ \bibnamefont
  {Setlur}},\ }\href@noop {} {\bibfield  {journal} {\bibinfo  {journal}
  {Physics Letters A}\ }\textbf {\bibinfo {volume} {383}},\ \bibinfo {pages}
  {3149} (\bibinfo {year} {2019}{\natexlab{b}})}\BibitemShut {NoStop}%
\bibitem [{\citenamefont {Stone}(1994)}]{stone1994bosonization}%
  \BibitemOpen
  \bibfield  {author} {\bibinfo {author} {\bibfnamefont {M.}~\bibnamefont
  {Stone}},\ }\href@noop {} {\emph {\bibinfo {title} {Bosonization}}}\
  (\bibinfo  {publisher} {World Scientific},\ \bibinfo {year}
  {1994})\BibitemShut {NoStop}%
\bibitem [{\citenamefont {Aristov}\ and\ \citenamefont
  {W{\"o}lfle}(2009)}]{aristov2009conductance}%
  \BibitemOpen
  \bibfield  {author} {\bibinfo {author} {\bibfnamefont {D.}~\bibnamefont
  {Aristov}}\ and\ \bibinfo {author} {\bibfnamefont {P.}~\bibnamefont
  {W{\"o}lfle}},\ }\href@noop {} {\bibfield  {journal} {\bibinfo  {journal}
  {Physical Review B}\ }\textbf {\bibinfo {volume} {80}},\ \bibinfo {pages}
  {045109} (\bibinfo {year} {2009})}\BibitemShut {NoStop}%
\bibitem [{\citenamefont {Danny~Babu}\ \emph {et~al.}(2020)\citenamefont
  {Danny~Babu}, \citenamefont {Das},\ and\ \citenamefont
  {Setlur}}]{danny2020density}%
  \BibitemOpen
  \bibfield  {author} {\bibinfo {author} {\bibfnamefont {N.}~\bibnamefont
  {Danny~Babu}}, \bibinfo {author} {\bibfnamefont {J.~P.}\ \bibnamefont {Das}},
  \ and\ \bibinfo {author} {\bibfnamefont {G.~S.}\ \bibnamefont {Setlur}},\
  }\href@noop {} {\bibfield  {journal} {\bibinfo  {journal} {Physica Scripta}\
  }\textbf {\bibinfo {volume} {99}},\ \bibinfo {pages} {105944} (\bibinfo
  {year} {2020})}\BibitemShut {NoStop}%
\bibitem [{\citenamefont {Das}\ \emph {et~al.}(2019)\citenamefont {Das},
  \citenamefont {Chowdhury},\ and\ \citenamefont {Setlur}}]{das2019nonchiral}%
  \BibitemOpen
  \bibfield  {author} {\bibinfo {author} {\bibfnamefont {J.~P.}\ \bibnamefont
  {Das}}, \bibinfo {author} {\bibfnamefont {C.}~\bibnamefont {Chowdhury}}, \
  and\ \bibinfo {author} {\bibfnamefont {G.~S.}\ \bibnamefont {Setlur}},\
  }\href@noop {} {\bibfield  {journal} {\bibinfo  {journal} {Theoretical and
  Mathematical Physics}\ }\textbf {\bibinfo {volume} {199}},\ \bibinfo {pages}
  {736} (\bibinfo {year} {2019})}\BibitemShut {NoStop}%
\end{thebibliography}%
\normalsize

\end{document}